\documentstyle[aps,prl,floats,twocolumn,epsf]{revtex}

\begin{document}


\wideabs{
\title{Coulomb Drag in the Extreme Quantum Limit}

\author{M.~P. Lilly$^1$, J.~P. Eisenstein$^1$, L.~N. Pfeiffer$^2$, and K. W. West$^2$}
\address{$^1$Condensed Matter Physics, Caltech, Pasadena CA 91125 \\
         $^2$Bell Laboratories, Lucent Technologies, Murray Hill, NJ 07974}

\maketitle

\begin{abstract}
Coulomb drag resulting from interlayer electron-electron scattering in
double layer 2D electron systems at high magnetic field has been
measured. Within the lowest Landau level the observed drag resistance
exceeds its zero magnetic value by factors of typically 1000. At
half-filling of the lowest Landau level in each layer ($\nu = 1/2$) the data
suggest that our bilayer systems are much more strongly correlated than
recent theoretical models based on perturbatively coupled composite
fermion metals. 
\end{abstract}

}

Double layer two-dimensional electron systems (2DES) have been the
subject of intense recent interest, especially at high magnetic fields,
owing to the diversity of many-body phenomena they exhibit which are not
found in ordinary single layer systems. These new phenomena arise from
the interplay of the intra- and interlayer Coulomb interactions and the
tunneling amplitude in the system. A particularly interesting case
arises when the individual 2D layers are at Landau level filling
fraction $\nu = 1/2$. If the separation between the layers is large the system
behaves as two independent 2DES's, each of which is widely believed
to be a Fermi liquid-like state of Chern-Simons composite particles. On
the other hand, when the layers are close together they behave as a
single system and exhibit a ferromagnetic quantized Hall state at total
Landau filling factor $\nu_{tot} = 1/2 + 1/2 = 1$. The nature of this remarkable
transition from two gapless Fermi liquids to a single gapped quantum
Hall phase is not well understood and remains a frontier topic
in the field\cite{review}.

The strength of the Coulomb interaction between electrons in opposite
layers is obviously a key ingredient of the physics. 
Recently a technique has been developed
which provides a simple way to directly obtain the interlayer electronic
momentum relaxation rate and thereby assess the strength of these
critical interactions. In this technique the frictional drag between the
two 2DES's is measured by observing the voltage which develops in one
layer when a current is driven through the other. This voltage, which
exists even though the two layers are electrically isolated, is directly
proportional to the interlayer momentum relaxation rate arising from the
scattering of electrons in one layer off those in the other. At zero
magnetic field drag studies have yielded a quantitative measure of the
Coulomb scattering rate
between electrons in the two layers\cite{zerodrag}, provided
evidence for momentum relaxation due to the exchange of
phonons\cite{phonons}, and revealed the predicted
plasmon enhancement of the drag\cite{plasmon.kf,plasmon.jpe,plasmon.nh,plasmon.tjg}.
Recent drag experiments
performed in magnetic fields large enough to induce the integer
quantized Hall effect have given evidence for the oscillatory screening
effects expected from Landau quantization\cite{iqhe.nb,iqhe.hr}.

In this paper we report the first Coulomb drag results from the extreme
quantum limit, focussing especially on the situation where in each 2D
layer the lowest Landau level is half-filled. Our measurements 
show that even though the
separation between the two layers is too large to yield the $\nu_{tot}=1$
quantized Hall state, the drag is considerably stronger
than recent theories predict for weakly coupled composite fermion (CF)
liquids. 

The samples employed in this work are modulation-doped GaAs/Al$_x$Ga$_{1-x}$As
double quantum wells grown by molecular beam epitaxy. They consist
of two 200~\AA\ wide GaAs quantum wells separated by a thin undoped
Al$_x$Ga$_{1-x}$As barrier layer. Although we have obtained qualitatively
similar results with samples having barrier thicknesses of 175~\AA\ and
100~\AA, most of the quantitative results presented here were obtained
using the latter. For this sample the barrier is pure AlAs (i.e. $x=1$).
Small differences in the sheet densities of the two wells were removed
using a Schottky gate electrode on the sample top surface. So balanced,
the 2D density in each quantum well was $N =1.38 \times 10^{11}$~cm$^{-2}$, and the
average low temperature mobility was $\mu \approx 1.7 \times 10^6$~cm$^2$/V s. For these
measurements simple Hall bar mesas were fabricated with
width $w=40$~$\mu$m and length $l=400$~$\mu$m. 

Separate electrical contacts to the individual 2D layers were
established via the ``selective depletion'' technique\cite{contact}. 
Drag measurements were performed by injecting a small 
drive current ($I \le 10$~nA @ 13~Hz) down the bar in one of the 2D
layers and recording the differential voltage $V_D$ appearing along the bar
in the other layer. Spurious signals arising from the capacitive
coupling of the two layers were minimized by keeping the measurement
frequency low and the common mode voltage of both layers small. Direct
measurements\cite{tunneling} of the tunneling resistances in our samples leave us
confident that tunneling is not influencing the drag data presented
here.

At zero magnetic field the drag data obtained from the 175~\AA\ sample is in
good agreement with that obtained earlier by 
Gramila {\it et al.}\cite{zerodrag,phonons}. As in this earlier data, the
temperature dependence of the drag scattering rate $\tau_D^{-1}$ (calculated from
the measured drag resistivity $\rho_D = -(w/l) V_D / I = (m/Ne^2) \tau_D^{-1}$ )
exhibits an anomaly around $T=2$K due to momentum relaxation resulting
from the exchange of phonons. In the 100~\AA\ sample, however, the zero
field drag is roughly twice as large as in the 175~\AA\ sample and the
phonon peak is thus less significant. This is expected since the
Coulombic part\cite{zerodrag,coulomb.part} 
of the drag is very sensitive to the separation between
the layers while the phonon part\cite{phonon.part} is not. At low temperatures the
Coulomb drag between two ordinary Fermi liquids in zero magnetic field
should vary quadratically with temperature: $\tau_D^{-1} \propto T^2$.
For the 100~\AA\ sample we find the ratio $\tau_D^{-1}/T^2$ to be constant to better
than 15\% over the temperature range $2 < T < 8$~K and on average equal to
$\tau_D^{-1}/T^2 \approx 4 \times 10^6$~s$^{-1}$K$^{-2}$.

On applying a magnetic field perpendicular to the 2D planes we find, in
agreement with earlier work\cite{iqhe.hr}, that the magnitude of the drag between the
layers increases dramatically. As at zero field, the {\it sign} of the induced
drag voltage is opposite to that of the resistive voltage drop in the
current-carrying layer. This is consistent with a momentum transfer
mechanism which sweeps carriers along in the dragged layer in the same
direction as those in the drive layer. We have also observed that the
magnitude of the induced voltage is unchanged when the roles of the
drive and dragged quantum wells are interchanged.

Fig.~1 compares the drag resistivity $\rho_D$ to the ordinary
longitudinal resistivity $\rho_{xx}$ of one of the 2D layers at $T = 0.6$~K and
magnetic fields $B > 5$~T. 
\begin{figure}
\begin{center}
\epsfxsize=3.3in
\epsffile[45 243 554 513]{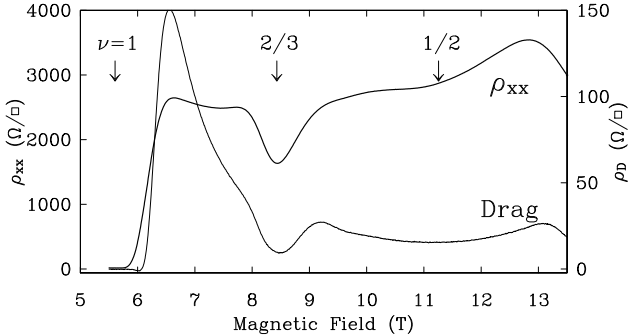}
\end{center}
\caption[figure 1]{Comparison of $\rho_{xx}$ to $\rho_D$ at $T = 0.6$~K.}
\end{figure}
(These data, and all that follow,
were obtained using the 100~\AA\ AlAs barrier sample.) Over the 
field range shown the (per layer) Landau level filling factor $\nu = hN/eB < 1$. 
Qualitatively, the two traces are quite similar: Where 
$\rho_{xx}$ exhibits a minimum due to a quantized Hall effect (e.g.
$\nu = 1$ and 2/3), so does $\rho_D$. This is not surprising since both
depend on the density of states
available for scattering. If a gap develops in the energy spectrum then
both electron-impurity and electron-electron scattering will be
suppressed and so therefore will be $\rho_{xx}$ and $\rho_D$. Note also that there is
no evidence in the data of Fig. 1 for any QHE at {\it total} filling factor
$\nu_{tot} = 1/2 + 1/2 =1$. In the present samples the layer spacing is too large
to support this intriguing QH state\cite{review}.

The similarity between the ordinary and drag resistivities disappears
when their temperature dependences are examined. As is well
known, $\rho_{xx}$ exhibits a strong variation with
temperature only in the vicinity of quantized Hall states. At non-QHE
filling fractions, like $\nu = 1/2$, $\rho_{xx}$ is only weakly dependent on
temperature. Indeed, for our sample $\rho_{xx}$ at $\nu = 1/2$ varies
(increases) by only 6\% when the temperature is reduced from $T = 4$~K to
0.2~K. In contrast, over the same temperature range $\rho_D$ at
this filling factor decreases by about a factor of 40.

Fig. 2 shows the temperature dependence of 
$\rho_D$ at $B = 11.45$~T, corresponding to filling factor $\nu = 1/2$ in each
layer. 
\begin{figure}
\begin{center}
\epsfxsize=3.3in
\epsffile[45 234 549 558]{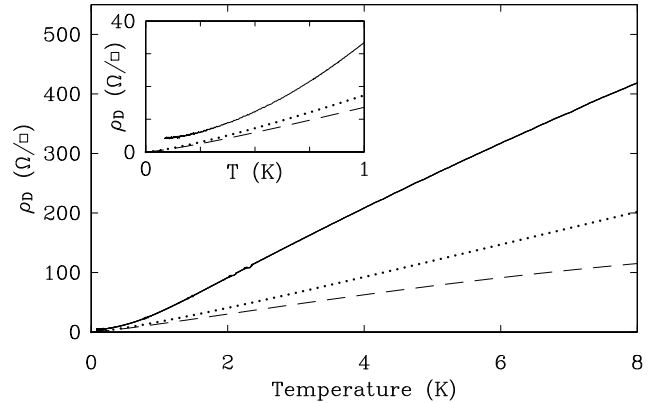}
\end{center}
\caption[figure 2]{Temperature dependence of $\rho_D$ at $\nu = 1/2$ (solid line).  
The broken lines are calculations\cite{cf.iu,calc} of $\rho_D$ assuming two different
CF effective masses (dotted, $m^* = 12 m_b$, dashed, $m^* = 4 m_b$, where $m_b$ is
the GaAs band mass).}
\end{figure}
These data illustrate the huge enhancement of frictional effects
on going to high field: At $B = 0$ the drag at $T = 4$~K is only 
$\rho_D \approx 0.1 \Omega / \Box$
while at $\nu = 1/2$ it is some 2000 times larger. 
In addition to this the drag at high field exhibits a different
temperature dependence than at $B = 0$. In particular, in neither the 100~\AA\
nor 175~\AA\ barrier sample does the $\nu = 1/2$ data show any evidence of a
``phonon peak'' in $\tau_D^{-1}/T^2$ similar to that seen at $B = 0$. While 
phonon-mediated drag at high magnetic field has not been
theoretically studied, it seems plausible that a such a peak might be
observed when the thermal phonon wavevector matches $2k_F$ for the
CF's. The absence of such a feature, plus the vast
enhancement of the drag itself, suggests to us that phonons are
relatively unimportant in our samples.

Recent theoretical work\cite{cf.ybk,cf.ss,cf.iu} has concluded that at 
low temperatures the
Coulomb drag resistivity between two clean 2DES's, each in
the $\nu = 1/2$ CF metallic state, ought to scale with temperature as $T^{4/3}$ and
{\it not} as $T^2$ as expected for simple Fermi liquids. This result has been
traced to the unusual wavevector and frequency dependence of the
conductivity $\sigma_{xx}$ of the CF liquids. In addition, 
Ussishkin and Stern\cite{cf.iu}
(US) have calculated the magnitude of the drag, its
dependence upon density, and the leading corrections to the $T^{4/3}$ law 
within a model of two perturbatively coupled 2DES's. In agreement
with experiment, they find the drag at high magnetic field to be far
larger than at $B = 0$; they attribute this to the very slow relaxation of
charge density fluctuations characteristic of the extreme quantum limit.
There is not, however, good quantitative agreement between their theory
and our experimental results. The two broken lines in Fig.2 show the
results of the US calculation for two reasonable choices of the
CF effective mass. The theory substantially
underestimates the observed drag\cite{calc}. 
US have also argued that disorder actually
reduces the drag and thus further degrades the comparison with
experiment. 

Predictions have also been made for how the drag should vary when
the densities of both layers are changed symmetrically (adjusting
B to maintain $\nu = 1/2$) and asymmetrically where, at fixed
B, the density of one layer is increased and the other equally
decreased\cite{cf.iu}. For the symmetric case theory gives $\rho_D \propto N^{-4/3}$ at low
temperatures. We find a significantly stronger
dependence. By combining a negative voltage ($V_{TG} = -0.4$~V) on the top
gate covering the drag mesa with a small dc bias voltage ($V_i = 15$~mV)
applied {\it between} the two 2D layers, we were able to reduce the density of
each layer from 1.38 to $1.08 \times 10^{11}$~cm$^{-2}$. This produced a roughly 70\%
increase in the drag whereas the change in $N^{-4/3}$ is only 39\%. For small
{\it asymmetric} density changes US find a quadratically increasing drag: 
$\Delta \rho_D / \rho_D = \beta ( \Delta N/\langle N \rangle)^2$ with $\Delta N$
the total density difference between the
layers, $\langle N \rangle$ the average single layer density, and $\beta = 7/48$. 
To produce such asymmetric density changes requires only the voltage $V_i$ applied
between the layers. Fig. 3 shows the results, at two representative
temperatures, of such experiments at $B = 11.45$~T where at $V_i = 0$ each 2D layer is 
at filling factor $\nu = 1/2$. 
\begin{figure}
\begin{center}
\epsfxsize=3.3in
\epsffile[63 243 567 567]{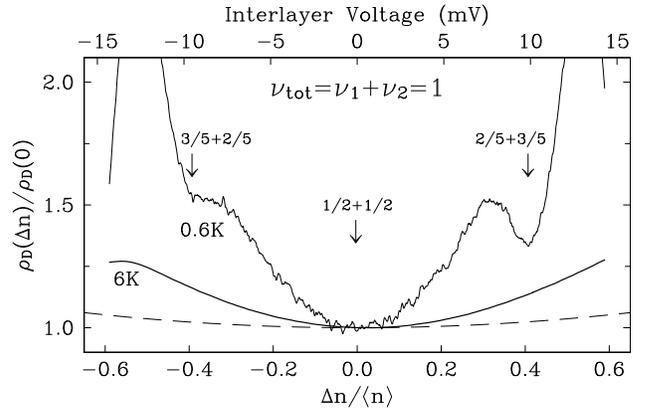}
\end{center}
\caption[figure 3]{Normalized dependence of drag on shifts of charge density from one 2D
layer to the other at $B = 11.45$~T.  Total Landau filling factor remains at 
$\nu_{tot}= \nu_1 + \nu_2 = 1$.   Dashed line is theoretical prediction\cite{cf.iu}.}
\end{figure}
The data have been normalized by the value observed with $V_i = 0$. The $T = 0.6$~K
trace shows well-defined features\cite{features} around $V_i \approx \pm 10$~mV that can be
associated unambiguously with one layer being driven into the $\nu = 2/5$
fractional QH state while the other goes into the $\nu = 3/5$ state. These
features provide a simple calibration of the density shift $\Delta N / \langle N \rangle$ 
vs. $V_i$.
Qualitatively, the data support the theoretical prediction of
a quadratic increase of the drag with $\Delta N$. It is clear however, that the
observed strength of the effect is much larger than the theoretical
prediction (the dashed line in the figure). 
For the $T = 0.6$~K data a least
squares parabolic fit gives $\beta \approx 6$, some 40 times greater than the
theoretical value. 
We find $\beta$ to increase as the temperature is reduced;
from $T = 0.2$ to 6~K our results are well fit by $\beta^{-1} \approx 0.08+0.16$~T. 

Turning again to the temperature dependence of the drag at $\nu = 1/2$, we
note that while at all temperatures our data exhibit a sub-quadratic
variation with $T$, they do not provide compelling evidence for the
predicted $T^{4/3}$ dependence. This is not surprising since it is known\cite{cf.ss,cf.iu}
that corrections to this power law occur at both high and, owing to the
disorder in the sample, at low temperatures. More importantly,
our data exhibit unexpected
behavior in the low temperature limit. As the inset to Fig.~2 reveals,
the drag appears to {\it remain finite} as $T \rightarrow 0$. 
This peculiar result,
which is at odds with the conventional picture of drag as resulting from
inelastic scattering events for which the phase space vanishes at $T = 0$,
led us to re-examine our measurement scheme in search of spurious
effects. In particular, the possibility of capacitive and resistive
leakage effects was carefully considered. No dependence on the
measurement frequency or the precise grounding configuration of either
2D layer was found in the in-phase resistive drag signal. 
Capacitive coupling was likely responsible for the {\it quadrature} signal
which was generally present, but its small magnitude and linear
frequency dependence made it easy to discriminate against. Finally,
although a non-zero drag in the $T \rightarrow 0$ limit was observed over a range of
filling factor around $\nu = 1/2$, it was {\it not} a ubiquitous feature of our
results. Very much smaller low temperature drag resistances were
observed at $B = 0$, inside QH states, and, interestingly, at and around
$\nu = 3/2$. These observations suggest that simple circuital coupling effects
are not responsible for the residual drag effect.

If the 2D electrons are heated out of equilibrium with the system
thermometer then an apparently non-zero drag at $T = 0$ would result. To
examine this possibility we studied the dependence of the drag
resistance on excitation current. Above about 0.5~K no non-linear effects
were found at any magnetic field. Below this
temperature non-linearities do appear, but in a way that is unexpected and
not consistent with simple heating. Fig. 4 compares the drag
resistivity observed at $T = 0.3$~K using $I = 7.5$~nA and 1.5~nA excitation
currents. 
\begin{figure}
\begin{center}
\epsfxsize=3.3in
\epsffile[45 234 513 504]{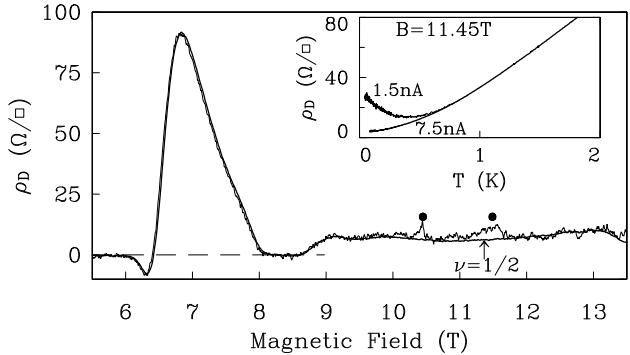}
\end{center}
\caption[figure 4]{Drag resistivity at $T = 0.3$~K at $I = 7.5$~nA (smooth
curve) and 1.5~nA (noisy curve).  Although the two curves are indistinguishable
below $B = 9.5$~T, the dots indicate fields around which the drag is significantly
non-linear.  Inset: Drag temperature dependence at $B = 11.45$~T 
for $I = 7.5$ and 1.5~nA.}
\end{figure}
Over most of the field range shown, these two currents yielded
essentially identical results. This is true even at
fields where the drag exhibits a specially large temperature dependence,
e.g. the flanks of the $\nu = 1$ and 2/3 QH minima. Surprisingly though,
non-linearities {\it are} evident 
in the general
vicinity of $\nu = 1/2$. At $B = 11.5$~T, for example, the drag is significantly
larger at $I = 1.5$~nA than it is at 7.5~nA. This is just the opposite of what
a simple heating model 
would imply.
As the temperature is reduced these non-linearities become stronger and
more widespread. At very low currents ($I < 0.5$~nA) a linear regime seems to
reappear but the poor signal-to-noise ratio at these currents leaves
this conclusion tentative. The inset to Fig. 4 shows the temperature
dependence of the drag at 11.45~T measured at 7.5 and 1.5~nA. It is clear that these
non-linearities can become quite large and, remarkably, that the drag at
small current can grow dramatically as the temperature is reduced. We
emphasize that these non-linear effects do not appear to be specific to
the $\nu = 1/2$ state but have so far been found throughout the range
$0.6 > \nu > 0.4$. While stable with time (for at least several days) it is
possible to alter their precise ``magneto-fingerprint'' by making large
changes in the experimental parameters (e.g. gate voltages, magnetic
field, temperature, etc.) Finally, we have even observed non-linearities
in which the small current drag is not only larger in magnitude but is
{\it of opposite sign} to that found using higher currents. 

In this paper we have reported Coulomb drag measurements on double layer
2D electron systems in the extreme quantum limit, focussing especially
on filling factor $\nu = 1/2$. For $T \stackrel{>}{_\sim} 0.5$~K our results,
while in qualitative agreement with recent theoretical models of weakly
coupled composite fermion liquids, point to significantly stronger
interlayer couplings than such models include. We have also observed
that the Coulomb drag does not always appear to vanish in the limit 
$T \rightarrow 0$
and that in the vicinity of the half-filled configuration remarkable
non-linear effects can appear. These effects, which recall charge
density wave depinning, are not understood. It is interesting to
speculate on whether they point to the existence, perhaps only in
isolated regions of the sample, of an unexpected bilayer ground 
state\cite{ground.state}.

We gratefully acknowledge useful discussions with N. Bonesteel, W.
Dietsche, S.M. Girvin, B.I. Halperin, Y.B. Kim, A.H. MacDonald, 
K. Yang, and especially A. Stern and the superb technical
assistance of S. Stryker. One of us (JPE) is also indebted to Bell
Laboratories, Lucent Technologies for supporting the initial stages of
this work.


\begin{references}
\bibitem{calc} To evaluate the theory we have taken the layer separation to be 300~\AA, 
the center-to-center distance between the quantum wells.
\bibitem{features} The different strength of these two features is likely due to different levels of disorder in the two layers.
\end{references}
\end{document}